# Arrays in Practice

## An Empirical Study of Array Access Patterns on the JVM

Beatrice Åkerblom[a]  and Elias Castegren[b]

a   Department of Computer and System Sciences, Stockholm University, Sweden
b   Department of Information Technology, Uppsala University, Sweden

**Abstract**   The array is a data structure used in a wide range of programs. Its compact storage and constant time random access makes it highly efficient, but arbitrary indexing complicates the analysis of code containing array accesses. Such analyses are important for compiler optimisations such as bounds check elimination. The aim of this work is to gain a better understanding of how arrays are used in real-world programs. While previous work has applied static analyses to understand how arrays are accessed and used, we take a dynamic approach. We empirically examine various characteristics of array usage by instrumenting programs to log all array accesses, allowing for analysis of array sizes, element types, from where arrays are accessed and to which extent sequences of array accesses form recognizable patterns. The programs in the study were collected from the Renaissance benchmark suite, all running on the Java Virtual Machine.

We account for characteristics displayed by the arrays investigated, finding that most arrays have a small size, are accessed by only one or two classes and by a single thread. On average over the benchmarks, 69.8 % of the access patterns consist of uncomplicated traversals. Most of the instrumented classes (over 95 %) do not use arrays directly at all. These results come from tracing data covering 3,803,043,390 array accesses made across 168,686 classes. While our analysis has only been applied to the Renaissance benchmark suite, the methodology can be applied to any program running on the Java Virtual Machine. This study, and the methodology in general, can inform future runtime implementations and compiler optimisations.

## The Art, Science, and Engineering of Programming

## 1 Introduction

Arrays are provided as a built-in data structure in most imperative programming languages. They are conceptually easy to understand and use, and their minimalistic functionality to collect and store more than one item in a sequence under the same name is useful in many programs. Arrays offer constant time random access and low storage overhead e.g., compared with a linked list where additional memory is needed to store the linking information. Arrays are used in a wide range of programs; from the simplest programs written by beginner programmers to programs making complex calculations on huge amounts of data.

At the same time as the array structure suggests a sequential access to the elements stored, the random access possibilities gives the programmer freedom to use any access order. This complicates reading and understanding code containing array accesses, especially when arrays are aliased or used in parallel programs. Static analysis of array accesses is an important prerequisite for many compiler optimisation techniques and program transformations, e.g., array bounds check elimination [16, 17], data dependence analysis [10, 14] and array access regions [12] for automatic parallelisation. This applies both to general characteristics such as array sizes and which parts of the arrays are actually accessed, but also to more intricate analysis of array accesses. This analysis could be as easy as collecting indices from the source code, but indices are generally represented by expressions, rather than literals. This means that exact indices are expensive, and in general even undecidable, to calculate statically. In theory, the complexity of array index calculations is unlimited and a general analysis must consider this. Previous work however indicates that the actual access patterns generated by complex index expressions turn out to be simple [12].

The motivation behind the work presented in this article is to continue the work made by Paek, Hoeflinger and Padua [12] on mapping out how array elements are actually accessed in practice and the degree of regularity of patterns that emerge. This will improve the foundations for further compiler enhancements.

This study was made on Java and Scala programs from the Renaissance benchmark suite [13] running on the Java Virtual Machine (JVM). The programs in the benchmark suite were executed and all array accesses made from within the programs were recorded. This approach permits the study of all sorts of patterns, ranging from simple traversal patterns controlled by e.g., a for-loop, to patterns resulting from accesses made in more complicated patterns and even random order.

We make the following contributions:

- We present a method for capturing general characteristics of arrays together with all accesses made to them running on the JVM (Section 2).
- We present a strategy for detecting patterns and sub-patterns in traces of array accesses (Section 2).
- We show that, on average, 69.8 % of the array access patterns for each benchmark consist entirely of combinations of uncomplicated sequential traversals and constant accesses (Section 3).

- We show that 53.8 % of all array accesses are made within identifiable uncomplicated sequential traversals or constant sequences (Section 3).

To understand array usage in "real-world" programs, we ask the following questions:

Q1 *What are the characteristics of arrays created and used in "real-world" programs?* This includes looking at metrics such as array sizes, what data types are stored in arrays and how large parts of arrays that are actually accessed.

Q2 *From where are arrays accessed?* This incorporates collecting data about where accesses are made from, tracing both from what classes the accesses are made and the identity of all threads that access an array.

Q3 *Are arrays accessed in regular patterns?* Answering this question requires studying the distribution of all accesses to individual arrays to discover emerging patterns, comparing the access distributions between arrays to find arrays that have been accessed in the same way, but also to find arrays that have been accessed in a unique way, and measuring the proportion of the arrays where arrays are made in regular traversals.

## 2 Method

The work presented in this article is the result of an empirical cross-sectional study: the data collection was performed once and data was collected from a set of representative, real-world programs. No experiments were performed and no variables in the environment of the running programs were manipulated to discover dependencies. The results of the study are therefore descriptive.

We chose a dynamic approach over a static analysis since it lets us produce an exact trace of all array accesses without approximations or abstractions. Since these traces depend on the specific execution paths taken by a program, the exact array access patterns may differ between runs of the same program.

### 2.1 Programs Studied

This study was performed on the 25 programs (7 Java and 18 Scala programs) included in the Renaissance benchmark suite [13]. Renaissance is an open-source project containing a large collection of programs implemented in Java and Scala, intended for testing "JIT compilers, garbage collectors, profilers, analyzers and other tools" [13]. The programs included were selected for their use of modern concurrency and parallelism models and using real-world workloads [13].

A short description of the programs in the Renaissance benchmark suite can be found in Appendix E.

### 2.2 Instrumenting the Programs

The Renaissance benchmark suite consists of programs written using Java and Scala, all running on the JVM. The programs are included in the benchmark suite as jar-files containing Java class-files. To make the classes output run-time information about array accesses, we instrumented 98.4 % of the class-files in the jar-files using the ASM framework [7]. More details can be found in Section 6 and in Appendix B.

In Java bytecode there are 16 bytecode instructions that deal with creation and access to arrays (see Appendix A for details). The bytecode instruction used depends on whether a value should be loaded or stored and on the type of the value.

To capture the information about array accesses from running Java bytecode, we first created a Java class containing methods that, when called, log data to a text file. Each of the 16 Java bytecode instructions that load values from or store values in arrays were associated with a dedicated method to log information about each use of any of these bytecode instructions.

Using the ASM framework [6] we built a tool to traverse Java bytecode classes, locate all uses of the 16 bytecode instructions of interest. When one of the traced instructions were located, new instructions were inserted to make calls to the right method in our logging class. A consequence of this approach is that information will be collected about all array accesses made within the instrumented classes but not for array accesses made in Java's standard library, e.g., when an array is used as an underlying data structure in Java's own ArrayList class.

Our tool was run on all class files in Renaissance except for the Renaissance framework itself and the ASM framework (which is also included in Renaissance).

## 2.3 Data Collection

The data collection was made using a dynamic, trace-based approach where the classes were instrumented to output information about array operations as they were performed. This means that the identity of the array and the specific index and whether the operation was read or write is known when the logging is made. Since the identity of the array used by the JVM is its hash code, which is is represented by 32 bits, it cannot be guaranteed to be unique. The risk of hash collisions will increase with the number of objects and in the data collected the risk will increase with the number of arrays of a certain type. Each access will be made from a specific line in a specific class running in a specific thread, and the identity of all of these is also known when logging. The line number and class name was collected from the running class file and the thread identity was collected by inserting a call to `Thread.currentThread().getId()`.

In the log files were collected: the identity of the array (the type of the array followed by its hash code), the accessed index, the mode (read or write), the array's length, the line number in the source code, the hash code of the class and the identity of the executing thread. Most parts of the information needed for this logging will be received as arguments by the logging method while other parts are captured from within the method when executed, e.g., thread identity. All fully qualified class names and their hash codes were also saved separately.

▪ **Table 1** Example of information stored about the accesses to an individual array.

| Lines from access data file | | | | | Explanation |
|---|---|---|---|---|---|
| [Lscala.collection.mutable.HashMap$Node;@7d8995e 16 12 | | | | | (first line) |
| r | 0 | 1 | 86  | F4C3E8A7 | **First line:** |
| r | 0 | 1 | 224 | F4C3E8A7 | Col 1: array id |
| w | 0 | 1 | 226 | F4C3E8A7 | Col 2: array length |
| r | 1 | 1 | 86  | F4C3E8A7 | Col 3: number of accesses |
| r | 1 | 1 | 224 | F4C3E8A7 | |
| w | 1 | 1 | 226 | F4C3E8A7 | **All the other lines:** |
| r | 2 | 1 | 86  | F4C3E8A7 | Col 1: mode (read or write) |
| r | 2 | 1 | 224 | F4C3E8A7 | Col 2: index accessed |
| w | 2 | 1 | 226 | F4C3E8A7 | Col 3: thread id |
| r | 3 | 1 | 86  | F4C3E8A7 | Col 4: line number from source |
| r | 3 | 1 | 224 | F4C3E8A7 | Col 5: class id |
| w | 3 | 1 | 226 | F4C3E8A7 | |

After instrumentation, all 25 benchmarks in the Renaissance suite were run according to the benchmark suite instructions using Java version 18.[1]

An example of stored information about the accesses observed for an individual array can be found in Table 1. This example lists all accesses made to an array of hash-map nodes (revealed by the fact that the array id starts with "[Lscala.collection.mutable.HashMap$Node"), where the array length was 16. The number of accesses registered is 12 and there were both reads and writes made to the indices 0, 1, 2 and 3, all made by thread number 1. The accesses were made from lines 86, 224 and 226 in the code and all accesses were made from the same class. All class names and their hash codes were saved during the instrumentation process. A look-up informs us that the class in question is scala/collection/mutable/HashMap.class.

### 2.4 Identifying Access Patterns

The access patterns in this analysis are sequences of tuples of index, mode (read or write), and thread identity.

Since the access pattern only considers the order in which indices were accessed (together with the mode and thread id), accesses to arrays of different types and length can exhibit the same pattern. The sequence of e.g., read accesses to index 1, 3, 5, and 7 may take place in any array with a length of at least length 8. Since access was made to only a subset of the elements in many arrays (see Section 3.2), it is possible that arrays of various lengths may have been accessed using an identical pattern.

Concerning the thread identity, the actual identity number is of less significance than the number of thread identities registered as operating on the array and the order in which they appeared. To enable comparisons based on these requirements, the thread identity is normalised to always start with number 1 for the first thread observed in the pattern, number 2 for the second one and so on. After normalising the thread numbers, all the access patterns for each benchmark are compared using their

---

[1] java version "18" 2022-03-22 Java(TM) SE Runtime Environment (build 18+36-2087) Java HotSpot(TM) 64-Bit Server VM (build 18+36-2087, mixed mode, sharing).

**Arrays in Practice**

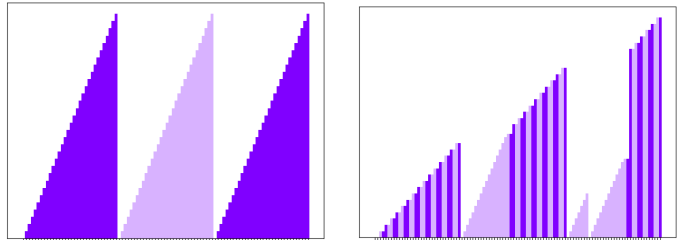

**Figure 1** Two access patterns. The first has three increasing traversals of step size 1, first only writes, then only reads and then only writes. The second has four increasing traversals of varying increments in 2 and 4, with reads and writes interleaved.

hash identity and only the hash-unique ones (per benchmark) are saved together with the ID of all arrays that have been accessed according to that pattern.

The same access pattern may reoccur in more than one of the benchmarks, but due to the large difference in number of arrays and access patterns produced by the benchmarks, we treat the access patterns of each benchmark separately.

## 2.5 Access Pattern Sequencing

We started out by plotting and ocularly inspecting all access patterns from the benchmarks with fewer than 300 patterns to search for common shapes. After that we first arbitrarily plotted 1.500 access patterns from a few more benchmarks and repeated the ocular inspection. In both cases the same shapes seemed to be most common ones.

As a next step in the continued analysis, we sampled 500 access patterns from each of the benchmarks and plotted them. ocular A majority of the plotted access patterns exhibited repetitive behaviour such as repeated accesses to the same index and traversals where the index is increasing or decreasing in different ways, e.g., accesses to index 0, 1, 2,... or 0, 0, 2, 2, 4, 4,....

A large proportion of the plotted access patterns consists of more than one traversal or other sub-pattern. Examples of access patterns consisting of sub-patterns can be seen in Figure 1, where the x axis displays the order of accesses and the y axis displays the indices accessed. The full colour represents write accesses and the more transparent colour shows read accesses. Both examples consist of sub-patterns. In example 1, we see three sub-patterns; traversals where the index is always incremented by 1. The first and third consists of write accesses only and the second consists of read accesses only. In example number two we see four sub-patterns, where the first, second and fourth are increasing traversals where the index is sometimes repeated, sometimes incremented by one and sometimes by another number all with different combinations of read and write operations. The third sub-pattern is a traversal where index is always incremented by 1.

By manual inspection of plotted access patterns, we identified a number of shapes in the patterns or sub-patterns.

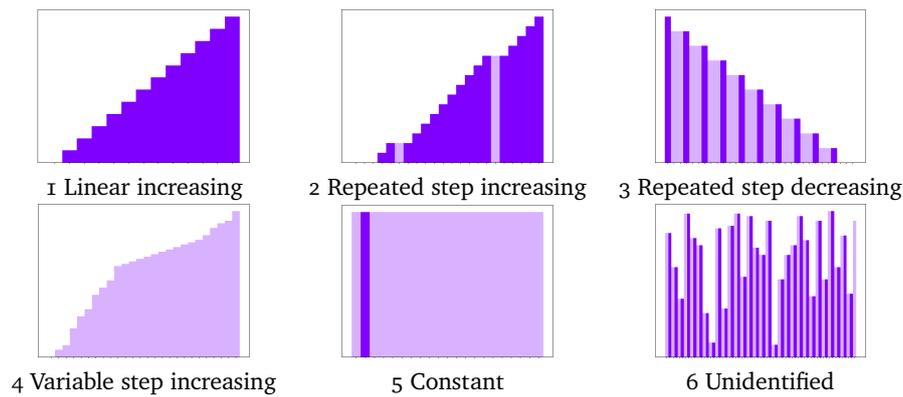

**Figure 2** Examples of array some access pattern shapes identified by manual inspection, and in example 6 an access pattern shape that is unidentified.

- *constant*, constant accesses, repeatedly accessing the same index
- *linear increasing/decreasing*, traversals where the index is linearly increasing or decreasing and where all steps between the indices have the same size
- *repeated step increasing/decreasing*, traversals where the index is increasing or decreasing and where at least one access revisits the previously accessed index
- *variable step increasing/decreasing*, traversals where the index is increasing or decreasing and where the step size between indices may vary

Plotted examples of the shapes in the list above can be seen in Figure 2. These example shapes were collected from the set of initially plotted access patterns. All the example shapes shown in Figure 2 consist entirely of one single shape, but most access patterns in our data consist of different combinations of the same or different shapes, like the examples in Figure 1, and sometimes together with unidentified parts.

All access patterns were then processed with scripts searching for the above shapes. If the entire access pattern did not fit into any shape when examined as a whole, the access pattern was split into slices and each slice in turn examined to see if they matched any of the shapes. The splits were made at all occurrences of the index 0 if there were more than one. If the index 0 did not occur repeatedly in the access pattern, the same approach was tried using index 1 (this heuristic is based on what we saw during manual inspection). As a consequence, all the "decreasing" patterns will either be a match for the entire access pattern, or it will always be preceded by an access to index 0 (or 1).

The presence of sub-patterns motivated a novel sequencing approach (inspired by gene sequencing) to capture the degree of regularity of the access patterns in terms of, e.g., repetitive behaviour and traversals. Sequencing in this case means translating the access pattern to a text representation or sequence of text representations that each indicate the position of an identified sub-pattern.

When an access pattern or a sub-pattern was identified as one of the shapes, a text representation was generated for the access pattern as a whole or each of its sub-patterns. The text representation for each access pattern or sub-pattern contains

**Arrays in Practice**

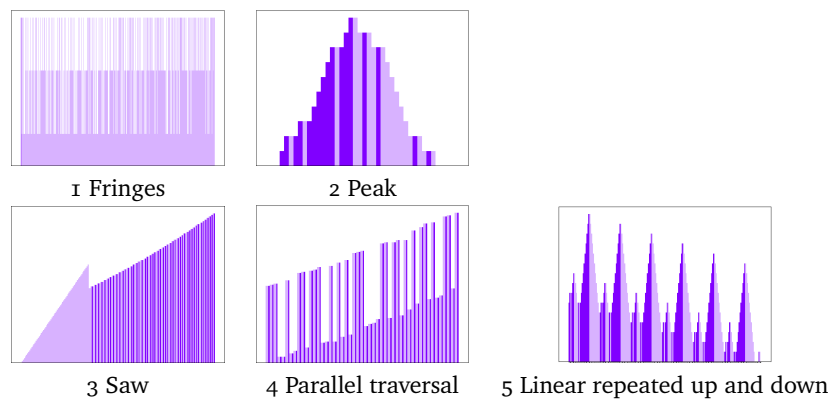

   1 Fringes    2 Peak

   3 Saw    4 Parallel traversal    5 Linear repeated up and down

■ **Figure 3** Examples of access patterns identified and searched for in the second round of the sequencing process. The x axis shows the time, the y axis shows the indices accessed. The full colour represents write accesses and the more transparent coulour represents read accesses.

information about what shape had been identified (if any), if the pattern was read-only, write-only or both read-and-write, the length of the pattern or sub-pattern, and the number of threads seen in the pattern. All patterns or sub-patterns that did not fit any of the shapes searched for were labeled as "unidentified". An example of an "unidentified" pattern can be found in example 6 in Figure 2.

As an example of the resulting text representation, the left access pattern in Figure 1 would appear as 0: |SLi w 1 42|SLi r 1 42|SLi w 1 42|. The initial 0 indicates the lowest index accessed. The three slices, separated by |, are all "strictly-linear-increasing" (SLi), split by accesses to the indicated lowest index. The first and last slice is write-only (w), while the middle is read-only (r). Each slice is accessed by a single thread (the 1) and the pattern is 42 accesses long.

After running the first round of the sequencing process, we repeated plotting 500 access patterns from each of the benchmarks, now solely from the access patterns that remained entirely unencoded. These access patterns were manually inspected in order to identify additional common shapes to search for in the next round of sequencing. In this process we identified five new common shapes:

- *fringes*, alternately accessing at a small number of indices
- *peaks*, start with a forward traversal and continue with a backward traversal
- *saws*, start with a forward traversal and continue with a forward traversal that restarts at a higher index than the start index of the initial traversal and where the length of the traversals are 2 or longer
- *parallel traversals*, traversals that are interleaved but access indices in two different, possibly overlapping, areas in the array
- *strictly linear repeated up and down*, small peaks, often on an increasing or decreasing base-line

Figure 3 contains plotted examples that visualise the five shapes above.

▪ **Table 2** The benchmarks, whether they were aborted when run or not, the number of accesses and the number of arrays identified.

| Benchmark name | Aborted | #accesses | #arrays | Benchmark name | Aborted | #accesses | #arrays |
| --- | --- | --- | --- | --- | --- | --- | --- |
| akka-uct | ☐ | 24,768,538 | 2,237,744 | movie-lens | ☑ | 271,149,972 | 1,073,186 |
| als | ☑ | 202,282,240 | 3,584,595 | naive-bayes | ☑ | 236,520,250 | 8,776,056 |
| chi-squared | ☐ | 210,498,191 | 6,279,502 | neo4j-analytics | ☑ | 219,865,850 | 16,095,319 |
| db-shootout | ☑ | 267,113,500 | 17,553,706 | page-rank | ☑ | 219,016,275 | 2,071,178 |
| dec-tree | ☐ | 249,910,766 | 2,434,128 | par-mnemonics | ☐ | 162,587 | 162,457 |
| dotty | ☐ | 35,874,111 | 266,933 | philosophers | ☑ | 215,502,508 | 4,735,505 |
| finagle-chirper | ☐ | 144,030,720 | 3,103,916 | reactors | ☑ | 164,514,805 | 14,356,462 |
| finagle-http | ☐ | 72,248,749 | 3,047,301 | rx-scrabble | ☐ | 2,103,071 | 230,923 |
| fj-kmeans | ☑ | 212,563,134 | 513,522 | scala-doku | ☑ | 235,149,442 | 442,552 |
| future-genetic | ☑ | 198,282,816 | 884,256 | scala-kmeans | ☐ | 57,664,333 | 65,324 |
| gauss-mix | ☑ | 288,559,562 | 1,381,705 | scala-stm-bench | ☐ | 63,067,456 | 3,379,128 |
| log-regression | ☑ | 200,149,810 | 189,232 | scrabble | ☐ | 11,882,117 | 620,493 |
| mnemonics | ☐ | 162,587 | 162,385 | | | | |

After running the second round of the sequencing process, we once again repeated plotting 500 access patterns from each of the benchmarks from the access patterns that still remained entirely unencoded. These access patterns were again manually inspected in order to identify additional common shapes. The manual inspection could identify some pattern-like structures in the plots but they were too irregular to classify with certainty by our analysis.

Many of the identified shapes described above are similar in their structure and many of the patterns that can be found in the data are borderline cases. The focus in the analysis is therefore to avoid inaccurate categorisation rather than to maximise the number of access patterns with a full coverage sequence.

# 3 Results

The programs run by the Renaissance benchmark suite consist of 171,481 classes and we successfully instrumented 168,686 (98.4%) of these. The identity of the uninstrumented classes and which benchmarks they belong to can be found in Appendix B. Some further notes about the uninstrumented classes can be found in Section 6.

## 3.1 Data Collection

Out of all the benchmarks, 12 were executed to finish while the 13 remaining were aborted during execution after they had generated more than 10 GB of array access information. Table 2 summarises information about the data collection. benchmarks resulted in 25 text In total 3,803,043,390 array accesses were registered in 93,647,508 arrays and the highest number of accesses made (288,559,562) were found in gauss-mix. The average benchmark contains 152,121,736 accesses to 3,745,900 arrays.

The observed array accesses were made from 24,690 call-sites in 5,843 classes out of the 168,686 that were instrumented. This means that arrays are used in only 3.5%

**Arrays in Practice**

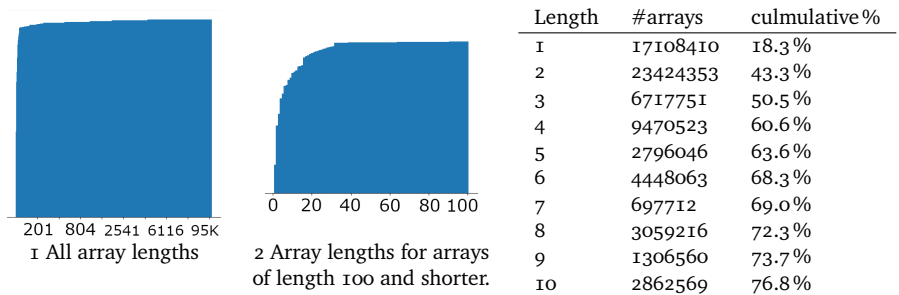

1 All array lengths

2 Array lengths for arrays of length 100 and shorter.

| Length | #arrays | culmulative% |
|---|---|---|
| 1 | 17108410 | 18.3% |
| 2 | 23424353 | 43.3% |
| 3 | 6717751 | 50.5% |
| 4 | 9470523 | 60.6% |
| 5 | 2796046 | 63.6% |
| 6 | 4448063 | 68.3% |
| 7 | 697712 | 69.0% |
| 8 | 3059216 | 72.3% |
| 9 | 1306560 | 73.7% |
| 10 | 2862569 | 76.8% |

■ **Figure 4** The cumulative number of arrays of different lengths. The **y axis** is the number of arrays of a certain length or shorter. The **x axis** is the lengths of the arrays. The left graph shows all the lengths of arrays. Note that the x axis is not proportional. The second graph is a zoomed in version with only arrays of lengths 100 and shorter. The table shows the numbers for arrays of length 10 and shorter.

of the classes in the Renaissance benchmark suite. The number of classes with array accesses is in reality somewhat lower, since the benchmarks were analysed separately and more than one benchmark may use the same library classes.

## 3.2 Characteristics of the Arrays in the Traces

Out of all arrays in our data, 71,891,203 (76.8%) contain 10 or fewer elements. Only few arrays (3.4%) are longer than 100 elements and very few (1.1%) exceed 1000 elements. There are arrays of all lengths up to 273. Above that, arrays of some lengths are not present. The number of absent possible lengths increases with the length. The longest array was used by the benchmark scala-stm-bench7 and it has capacity to store 2,359,299 elements. The distribution of all array lengths can be seen in Figure 4.

The ten most common lengths among the arrays we identified were 2, 1, 4, 3, 16, 6, 8, 10, 5, and 13 in that order. 43.3% of all arrays has capacity for 1 or 2 elements only. A zoomed in version of the cumulative distribution of array lengths showing only arrays of length 100 or shorter can be found in Figure 4, in example 2. The number of arrays identified with lengths 10 and shorter can also be found in the table of Figure 4, together with the cumulative share of all arrays.

Out of all the arrays, 75.3% were only accessed by code belonging to a single class. Further 23.5% of the arrays were accessed from two different classes, which means that only 1.2% of the arrays were accessed from three different classes or more. The array traced with accesses from most classes was accessed from 54 different classes in neo4j-analytics. The array is two elements long, contained elements of the type java.lang.Object and had 11,304,575 logged accesses.

For every access made, the length of the array was saved. For some arrays, in 20 out of the 25 benchmarks, the length logged differed between accesses. The number of affected arrays is low; this happens for only 0.22% of the arrays. A majority of the arrays (83.9%) for which the length changes in the logged data, it changes once, and most (85.5%) the change is made in one direction (growing or shrinking), even if some these lengths change size more than once. The length of some of the arrays

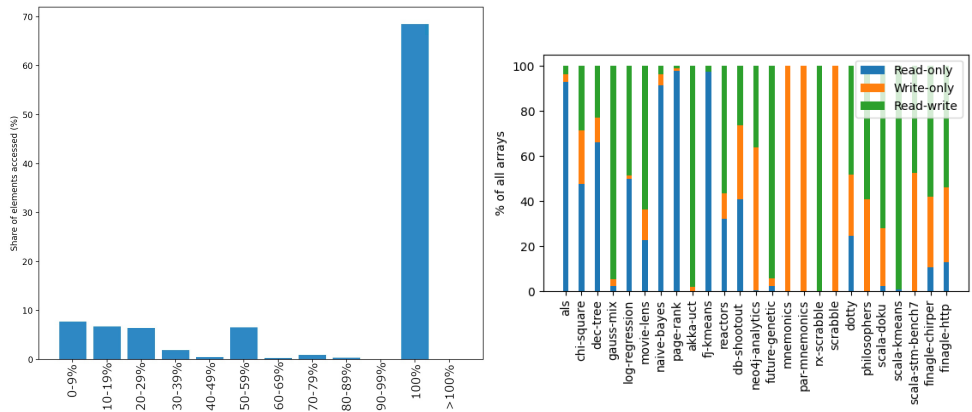

**(a)** Share of elements accessed in all arrays.    **(b)** Read/write-only arrays across benchmarks.

**Figure 5** The share of elements accessed, and distribution of read/write-only arrays.

that do change, (14.6 %) do however change more than once both increasing and decreasing.

All elements were accessed at least once in 68.5 % of the arrays. For 23.3 % of the arrays, less than 50 % of the indices were accessed and for 7.7 % of the arrays less than 10 % of the indices were accessed. In 0.08 % of the arrays, accesses were registered to indices outside the array's bounds. The distribution can be seen in Figure 5a.

For a large proportion, 33.6 % of the arrays, only read operations were logged. Another large proportion of the arrays, 28.9 %, were logged with only write operations. The read-only and write-only arrays are however not evenly distributed among the benchmarks. The distribution between read-only, write-only and read-write array can be seen in Figure 5b.

Some benchmarks contain mostly read-only arrays (e.g., page-rank: 97.8 %, fj-kmeans: 97.3 %, als: 93.0 %). Some other benchmarks contain mostly write-only arrays (e.g., mnemonics: 100.0 %, par-mnemonics: 100.0 %, scrabble: 100 %). All of these contain a few read-only arrays and read-write arrays. Another group of benchmarks contain mostly read-and-write arrays (e.g., rx-scrabble: 99.9 %, scala-kmeans: 99.1 %, akka-uct: 98.0 %, gauss-mix: 94.7 %).

All three categories (read-only, write-only and read-write arrays) were detected in the log files from all of the benchmarks, but in some cases the number is so low that the category is not visible in the chart.

The arrays in the study contained elements of 814 different types. The types included all of Java's primitive types, boolean, byte, char, double, float, int, long and short. There were also arrays containing different Java or Scala objects, other application-objects, and nested arrays of different dimensions and containing different objects. The type categories are summarised in Table 3. The most common element type was Java.lang.Object, 24.8 %. 49.2 % of the arrays contained a type from the Java standard library (including Java.lang.Object), 31.7 % of the arrays contained some primitive type, 8.8 % contained a type from the Scala standard library, 5.8 % contained some other object type, and 4.5 % contained nested arrays of different depth.

**Arrays in Practice**

There is no clear correlation between any of the element type categories mentioned above and the average length of the arrays in which they were found. The largest average length was found for java.util.concurrent.ConcurrentHashMap$Node and java.nio.channels.SelectionKey arrays with an average length of 1,032 and 1,024 respectively. The third largest average length, 220, was found in byte arrays, and the fourth, 139, in java.net.URL arrays. All other types were found in arrays with an average length below 100.

The most common nested array type is a two-dimensional array storing bytes, [[B. 98.3 % of the nested arrays had this type. The highest number of dimensions of nested arrays seen in a type was five and in the traces there were only two such arrays, both with the type[[[[[Ljava.lang.Object and length 32 at the outermost level. Nested arrays at levels four, three and two were also found, and the number of nested arrays increases as the nesting level decreases.

The types of the actual elements stored at the innermost level of the nested arrays includes all of Java's primitive types (except boolean), Java.lang.Object and a few other Java object types, some Scala object types, a few Apache Spark types and a couple of other object types (from Antlr, Neo4j, Breeze, Cafesat and Guava).

Even though most of the programs included in the Renaissance benchmark suite were selected for their use of parallel constructs, most arrays in these programs are used thread-locally; only 2.8 % (2.664.128) of the arrays were accessed by more than one thread. In four of the benchmarks no arrays at all were accessed by more than one thread (db-shootoout, mnemonics, scala-doku and scala-kmeans). The highest observed number of threads accessing the same array was 81 (in akka-uct) and the average highest number of threads among the benchmarks 14. For many of the arrays that were used by more than one thread, the observed accesses were both reads and writes. In 6 of the benchmarks these constitute more than 95 % of the arrays (gauss-mix, reactors, future-genetic, rx-scrabble, dotty and scala-stm-bench7) and the average share of read-and-write accesses among the arrays accessed by more than one thread was 64 %. In three of the benchmarks where multi-threaded array accesses were observed, less than 1 % of these arrays were both read and written (log-regression, fj-kmeans and par-mnemonics). The average length of multi-threaded arrays (25.5) is slightly lower than the average length of single-threaded arrays (36.9).

■ **Table 3** Element types, the number of arrays containing elements of these types, their share of the total arrays, the average, max and min length for the traced arrays.

| Element type | #arrays | %arrays | Avg len | Max len | Min len |
|---|---|---|---|---|---|
| Java stdlib | 46048962 | 49,17 % | 53,77 | 2385494 | 1 |
| Primitive | 29680185 | 31,69 % | 49,73 | 4950622 | 1 |
| Scala stdlib | 8277537 | 8,84 % | 9,43 | 1048576 | 1 |
| Other objects | 5437352 | 5,81 % | 9,26 | 1048576 | 1 |
| Array | 4203472 | 4,49 % | 20,85 | 1944 | 1 |

■ **Table 4** The number of access patterns identified for all benchmarks and the number of arrays that could be represented by each access pattern, on average.

| Benchmark name | #identified access patterns | Average array/ access pattern | Benchmark name | #identified access patterns | Average array/ access pattern |
| --- | --- | --- | --- | --- | --- |
| akka-uct | 15,051 | 148.7 | movie-lens | 5,711 | 187.9 |
| als | 5,000 | 716.9 | naive-bayes | 7,428 | 1181.5 |
| chi-squared | 3,199 | 1963.0 | neo4j-analytics | 42,273 | 380.7 |
| db-shootout | 32,641 | 537.8 | page-rank | 2,076 | 997.7 |
| dec-tree | 16,801 | 144.9 | par-mnemonics | 17 | 9556.3 |
| dotty | 30,457 | 8.8 | philosophers | 2,295 | 2063.4 |
| finagle-chirper | 8,501 | 365.1 | reactors | 105,546 | 136.0 |
| finagle-http | 2,204 | 1382.6 | rx-scrabble | 122 | 1892.8 |
| fj-kmeans | 8,003 | 64.2 | scala-doku | 12,749 | 34.7 |
| future-genetic | 51,200 | 17.3 | scala-kmeans | 300 | 217.7 |
| gauss-mix | 2,844 | 485.8 | scala-stm-bench7 | 19,503 | 173.3 |
| log-regression | 1,572 | 120.4 | scrabble | 40 | 15512.3 |
| mnemonics | 13 | 12491.2 | | | |

### 3.3 Access Patterns

Repeatedly executing the same code using different arrays may result in an identical access pattern, but if the access order in any way depends on values in the elements stored, the access pattern resulting from the executing the same code will differ. On the other hand, code with widely different purposes may result in identical access patterns. Some access patterns are used for accessing elements in many arrays, while others are unique. By identifying access patterns and grouping arrays that were accessed using the same access pattern, the 93,647,508 identified arrays in our data could be represented by a significantly lower number of access patterns in all benchmarks. The number of access patterns for each benchmark and the average number of arrays represented by each access pattern can be found in Table 4. The benchmark where the access patterns could represent the highest number of arrays, on average, was scrabble, with 15512.3 arrays per access pattern. The lowest average was found in dotty, where only 8.8 arrays could be represented by each access pattern. The average for all benchmarks was that each access pattern could represent 2031.2 arrays.

In total, the data from the benchmarks contains 375,546 access patterns where, as mentioned above, duplicates from different benchmarks are included. 295,124 of these access patterns were only used for accessing elements from one array each (within that benchmark). The single access pattern used on the largest number of arrays was found in the benchmark naive-bayes. It represent the use of 7,720,175 different arrays, that is 88.0 % of all arrays registered for that benchmark. This pattern contains two read accesses, first to index 0 and then to index 1, both made from the thread numbered 1, [[r 0 1][r 1 1]]. The distribution of all arrays over the access patterns identified can be seen in Figure 6.

### 3.4 Access Pattern Sequences

**Access Pattern Sequencing–Round One**  In the first round of the access pattern sequencing, we searched all access patterns for the traversals and constant shapes

**Arrays in Practice**

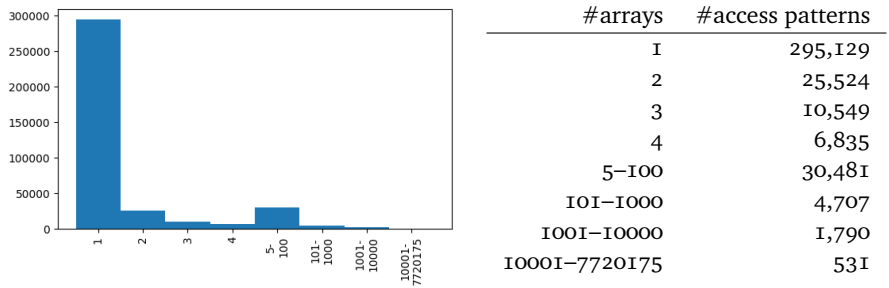

| #arrays | #access patterns |
|---|---|
| 1 | 295,129 |
| 2 | 25,524 |
| 3 | 10,549 |
| 4 | 6,835 |
| 5–100 | 30,481 |
| 101–1000 | 4,707 |
| 1001–10000 | 1,790 |
| 10001–7720175 | 531 |

■ **Figure 6** Each array has been accessed in a certain pattern. The access pattern may be unique or be shared by many arrays. The number of arrays sharing the same access patterns can be seen in the histogram and the table. 295,129 access patterns were unique for one single array, 25,524 access patterns were shared by two arrays etc. 30,481 access patterns were shared by between 5 and 100 arrays, 4,707 access patterns were shared by between 101 and 1,000 arrays and so on.

shown in Figure 2. All of these shapes were found in access patterns from all of the benchmarks. On average, for all benchmarks, 69.8 % of the access patterns could be entirely encoded as consisting of a sequence of these access pattern shapes. On average, for all benchmarks, 19.3 % of the access patterns contained parts that could be encoded, and 10.9 % of the access patterns were left completely unencoded.

The shares for each individual benchmark can be found in Figure 7. The share of access patterns that could be completely encoded ranges from 100 % (in mnemonics and par-mnemonics) to 18.8 % (in scala-stm-bench7). The share of access patterns that remained completely unencoded range from 0 % (in mnemonics and par-mnemonics) to 43.7 % (in scala-doku).

The access pattern level analysis does however not inform us about the actual share of all array accesses that could be identified as belonging to a certain shape. This is partially due to the fact that the actual size of the encoded part of the partially encoded access patterns may vary. Another factor is that the number of actual accesses made varies between access patterns; some access patterns are more common than others. The access pattern's lenght is yet another variable that needs to considered.

■ **Table 5** Distribution of array accesses over patterns.

**(a)** Distribution of accesses after round one.

| Access pattern shape | % of all accesses |
|---|---|
| Repeated step increasing | 30.8 % |
| Linear increasing | 23.0 % |
| Constant | 7.5 % |
| Linear decreasing | 7.5 % |
| Variable step increasing | 0.3 % |
| Repeated step decreasing | 0.1 % |
| Variable step decreasing | 0.0 % |
| Unidentified | 30.8 % |

**(b)** Distribution of accesses after round one.

| Access pattern shape | % of all accesses in |
|---|---|
| Saws | 8.1 % |
| Peaks | 4.1 % |
| Fringes | 0.7 % |
| Linear repeated up and down | 1.7 % |
| Parallel traversals | 0.3 % |
| Identified in round one | 69.2 % |
| Unidentified | 15.1 % |

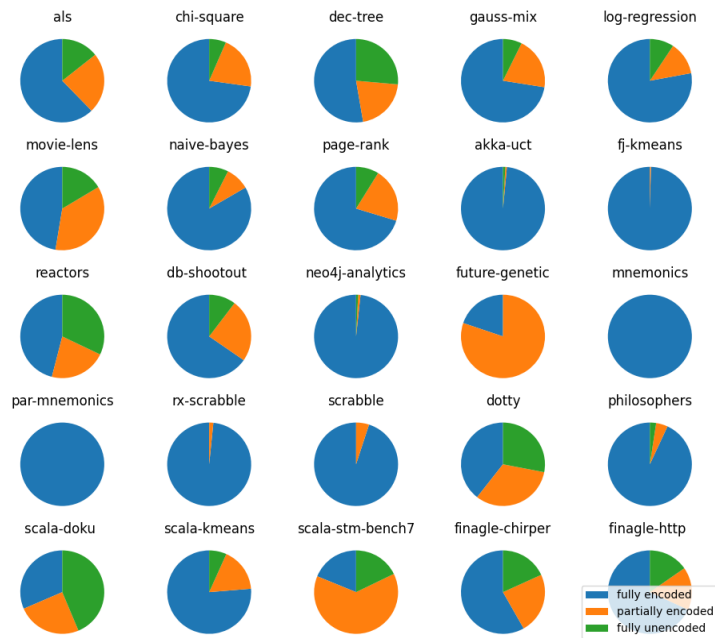

**Figure 7** The results of running the first sequencing process: the share of access patterns entirely (blue) and partially (orange) encoded, and fully unencoded (green).

At the individual access level, 69.2 % of all array accesses logged are part of an identified access pattern shape after the first round of encoding. The distribution of accesses between encoded parts of the access patterns (which can be partial) and unencoded parts of the access patterns for each benchmark can be found in Figure 8.

The share of accesses located in encoded parts of the access patterns, i.e. within an identified pattern shape, ranges from 100 % (in mnemonics and par-mnemonics) to 3.2 % (in rx-scrabble). The average share is 63.0 %.

The dominating shape in all the access patterns is the "repeated step increasing", which incorporates 30.8 % of all accesses. The second most common identified shape is the "linear increasing" which covers 23.0 % of all accesses. Table 5a shows the distribution of individual accesses over all the identified access pattern shapes collected from all benchmarks.

**Access Pattern Sequencing–Round Two** In the second round of the access pattern sequencing, we searched all access patterns for the same shapes as in round one, but also extended the search to include the shapes in Figure 3. The distribution between fully encoded, partially encoded and completely unencoded access patterns for each benchmark can be seen in Figure 9.

After the second round of encoding the share of access patterns that were entirely encoded had increased for all benchmarks but five. Two out of these had already been encoded to 100 % in the first round (mnemonics and par-mnemonics). The other three, (fj-kmeans, rx-scrabble and rx-scrabble), were all encoded to a high degree (between 95.0 % and 99.6 %) already in the first round.

**Arrays in Practice**

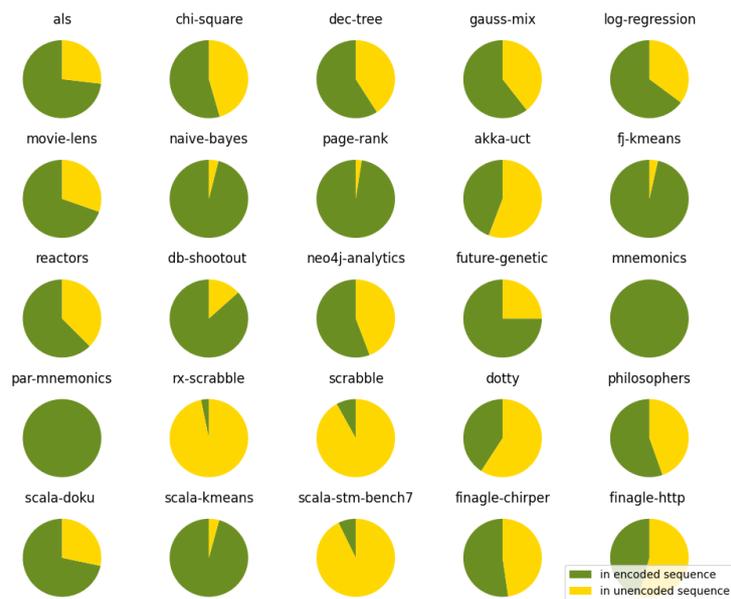

■ **Figure 8** The distribution of individual accesses identified as part of an identified access pattern shape (green) or part of an unidentified shape (yellow) after round one. sections represent the accesses

On average, for all benchmarks, 81.4 % of the access patterns could be entirely encoded as consisting of the extended set of access pattern shapes, compared to 69.8 % after the first round. The benchmark with the lowest share of fully encoded access patterns after the second round was still `scala-stm-bench7` (now 45.6 %, compared to 18.8 % after round one). The benchmark `scala-doku` still had the largest share of access patterns that remained completely unencoded, now 25.4 %, compared to 43.7 % after round one. On average, for all benchmarks, 7.2 % of the access patterns were completely unencoded (compared to 10.9 % after the first round).

Once again, we also divided the individual accesses into the shape categories that they were part of and the share of accesses that were identified as part of any of the original shapes stayed there. All accesses now identified as part of the new shapes were taken from the "unidentified" category from the first round of encoding.

After the second round, the share of individual accesses that were part of identified shapes had increased from 69.2 % to 76.2 % on average for the benchmarks. The distribution of individual accesses between encoded and unencoded (with an "unidentified" shape) parts of the access patterns for each benchmark after the second round of encoding can be found in Figure 10. The distribution of individual accesses over all access pattern shapes in the second round can be seen in Table 5b.

The share of accesses located in encoded parts of the access patterns ranges between 100 % (in `mnemonics` and `par-mnemonics`) and 7.9 % (in `scala-stm-bench7`).

The shapes from the first round were still dominating. The largest amount of newly identified accesses belonged to the "saw-tooth" shape, which incorporated 8.1 % of the accesses.

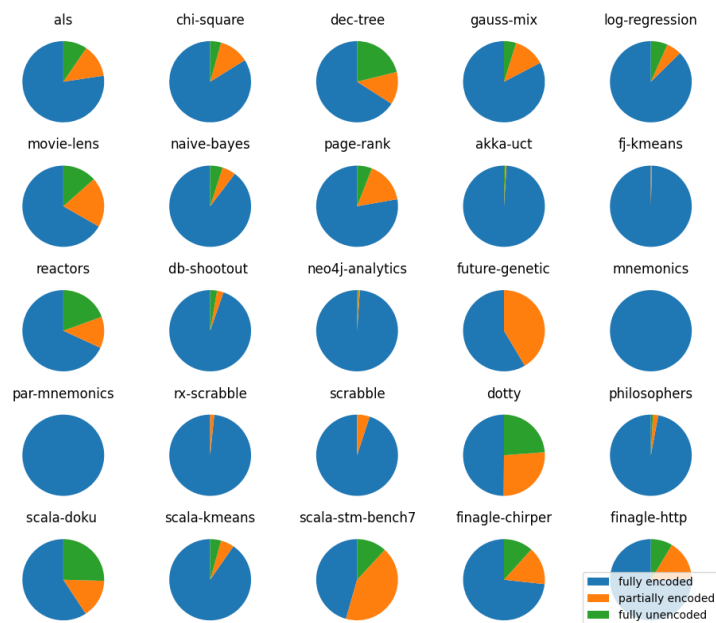

**Figure 9** The result after the second round of the sequencing process searching for an extended set of shapes.

## 4 Discussion

In this section we summarise and discuss the results presented in the previous section in relation to the research questions Q1-Q3 listed in Section 1.

### 4.1 Q1 What Are the Characteristics of Arrays Created and Used in "Real-World" Programs?

Most arrays are really short, containing 1–3 elements. In most of the arrays (68.5 %), all indices were accessed at least once. A large proportion of the arrays were logged with exclusively read or write operations.

For benchmarks with predominantly read-only arrays, a possible explanation is that the benchmarks write data to non-array data structures and then collect it in arrays through means not logged by our instrumentation, e.g., in standard library methods, like calling toArray() on a stream. This is discussed in more detail in Appendix C, with some illustrating examples. For the opposite case, with predominantly write-only arrays, a possible explanation may be that the end-goal of programs run in benchmarks is to measure performance. To this end, the generation of data (which is saved in an array) may be the purpose rather then the use of the generated data.

Another possible cause of the read-only and write-only arrays is that they are a result of the arrays being created or used through Java's reflecion library or other untracked libraries. This is discussed in more detail in Section 6.

A very small number of arrays, 0.22 %, were logged with different lengths for different accesses. This could be a result of that the array id stored is a combination of

**Arrays in Practice**

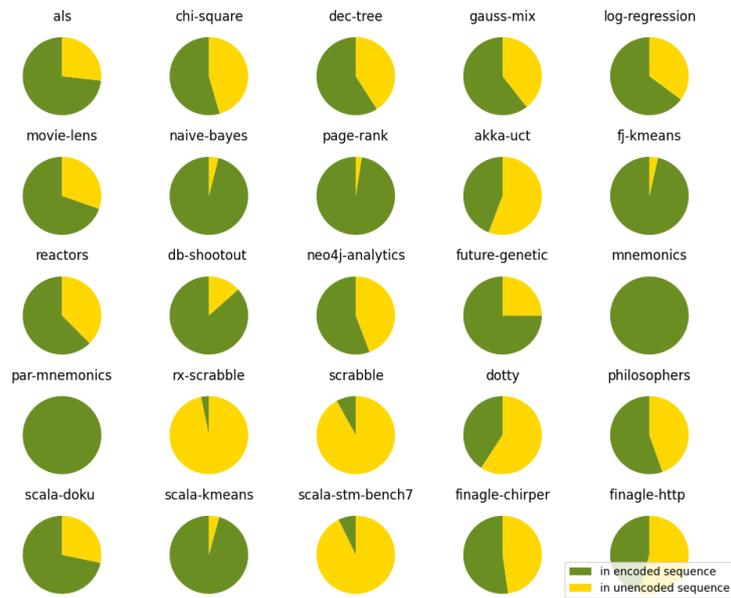

**Figure 10** The distribution of individual accesses identified as part of an identified access pattern shape (green) or part of an unidentified shape (yellow) after round two.

the array's type and the array's hash code, which cannot be guaranteed to be unique. A further discussion of this can be found in Section 6.

The types stored in the arrays were to a great extent (49.2 %) some object type from the Java standard library (including java.lang.Object). A possible cause of the large number of arrays with java.lang.Object elements is that the arrays are used through generics. Primitive types were also common element types (31.7 %).

### 4.2 Q2 From Where Are Arrays Accessed?

Most arrays are accessed by objects from one single class. This indicates that arrays are often not passed around between objects of different classes but they may still be accessed by more than one object of the same class. A vast majority of the arrays (97.2 %) were accessed from only one thread, even though most of the programs included in the Renaissance benchmark suite use parallel constructs.

### 4.3 Q3 Are Arrays Accessed in Regular Patterns?

Most access patterns (78.6 %) are unique and used for only one array. Nevertheless we have identified access patterns that are used for a very large number of arrays.

By searching for the shapes of simple traversals and constant accesses (as illustrated by Figure 2) in the access patterns we found that a majority (69.8 %) consisted entirely of these. The simple traversals and constant accesses include a majority of all accesses made to the arrays (69.2 %). The most common shape was the "repeated step increasing" (incorporating 30.8 % of all accesses), which could be the result of traversing an array, reading each element making some change to that element and writing it back to the array in the same position.

After the first round, we were able to identify a number of other common shapes (as illustrated by Figure 3) that we added to the analysis. With these new shapes included in the analysis, the number of access patterns that consisted entirely of the shapes searched for increased from 69.2 % to 81.4 %. The share of accesses made within an identified shape increased from 69.2 % to 76.2 % The most common of the added shapes was the "saw" which incorporates 8.1 % of all array accesses.

### 4.4 Possible Implications

While we have not evaluated any optimisations, our results support the efficiency of some existing optimisation techniques and suggest other viable strategies. The fact that most arrays are shorter than 10 elements and are accessed in predictable patterns suggests that loop unrolling and automatic vectorisation will often be efficient. Such short arrays will also fit in a cache line or two, meaning the access pattern will not matter for cache performance. For the really short arrays (three elements or shorter) one might even be able to replace the array with stack or instance variables, depending on the lifetime of the array, to avoid the garbage collection and memory overhead.

While we have not checked array accesses per object, the fact that many arrays are accessed from a single class suggests that they may often be encapsulated by an object, in which case short arrays may be embedded in the object as instance variables (again to avoid garbage collection overhead). Since two thirds of array accesses follow predictable patterns, this suggests that bounds checks can be elided for these, although statically detecting this is another problem. One could also imagine abstractions over arrays that *only* allow the patterns found in this paper, giving the same performance benefits as arrays with just the right amount of expressiveness for how they are used.

## 5 Related Research

This paper is related to work in two different areas; static access pattern analysis and trace-based analysis in general.

Statically calculating the exact array indices accessed in a program based on array subscripts is expensive and sometimes undecidable so a trade-off between efficiency and accuracy must be made. Program optimisation and program transformations at compile time requires array analyses to secure that accesses from different parts of the code will not interfere with each other [9, 11, 12, 14]. Much of the work in the area revolves around improving either efficiency or accuracy for array accesses analyses for different purposes. One interesting result from e.g., [12] and [3] is that access patterns are simple, even in cases where the indexing is made through complex expressions, a result in line with our findings that access patterns are simple to a great extent. Our study is aimed at the same problem, understanding the pattern of accesses to elements in arrays, but instead of performing static analysis on array accesses in source code we have traced the accesses when they occur in running programs.

**Arrays in Practice**

Trace-base analyses have been used to study many different aspects of running programs both for single studies e.g., [2, 4] or built into tools to facilitate analyses of e.g., parallelism in RoadRunner [8] and aliasing in Spencer[5]. Unlike RoadRunner and Spencer, our dynamic analysis focused solely on array usage.

While not about arrays, our work exists within the same sphere as Tempero, Yang and Noble's study of in Java [15], Åkerblom and Wrigstad's study of polymorphism in Python [2] and Brandauer and Wrigstad's study of aliasing and mutability in Java [4]. Like them, we are interested in empirically investigating how programs "in the wild" work and to what extent they correspond to preconceived notions of programming.

## 6 Threats to Validity

The programs included in the Renaissance benchmark suite may not be representative for all Java programs in terms of their use of arrays, but the programs included were selected, among other things, for their realistic workloads that reflects real-world program behaviour [13].

The array access patterns extracted from executing benchmarks may however display different characteristics than the array access patterns resulting from the normal program execution. Data may be produced and calculations performed but the outcome never used as it would have been if the program was actually used for its purpose. The programs, on the other hand, are real world programs and the part executed by the benchmark will also be part of that normal execution.

Of all the classes in the Renaissance benchmark suite, 98.4 % were instrumented and all array accesses made in these classes were logged. The ASM framework (223 classes in two separate versions included in db-shootout and neo4j-analytics respectively), which was used for the class-file instrumentation is included in Renaissance but was excluded from the instrumentation. The 2.556 classes belonging to the Spire framework were not instrumented since the jar files could not be unpacked. The Spire framework is included as part of apache-spark, which in total consists of 76,358 classes. 16 other classes failed instrumentation due to unresolved dependencies. We do not expect the 1.6 % of uninstrumented classes to affect the overall results in any significant way.

Our study is built on instrumentation of class-files, meaning any code generated at runtime will not be instrumented. All arrays that are created and used in code generated at runtime is consequently missing in our data. There are several ways to create arrays using Java's reflecion library; both by copying existing arrays and by creating new ones for which the element type is not statically known. 397, that is 0.2 %, of the 171,481 classes of the programs that are run by the Renaissance benchmark suite import java.lang.reflect.Array, which can be used for, e.g., creating arrays whose element types are not known at compile time and for reading from and writing to arrays. 1,871 classes, whereof 1,427 are present in our traces, use the native method System.arraycopy() which copies (parts of) arrays, and which is also not traced. This could account for some of the arrays that appear to only be read or written

The approach to instrument class-files has the consequence that Scala's library classes were instrumented, while Java's were not. To examine whether this skews the results, we separated all arrays that were used from classes within Scala's standard library from the arrays used from other classes and run the analysis separately on these, see Appendix D.1. The Scala-only access pattern analysis produces similar results to our original results. For the read-only and write-only arrays, the Scala only analysis has a higher average for read-only arrays and a lower average for the write-only arrays. Three out of four benchmarks that had over 90 % read-only arrays were Scala benchmarks, while all three benchmarks that had 100 % write-only arrays were Java benchmarks.

The data used to identify an array is the array object's hash code and the name of its element type. This identity cannot be guaranteed to be unique, especially not in programs where large quantities of arrays with the same length and element type are created. The JVM hash code is represented by 32 bits. If we assume a uniform distribution and apply the birthday paradox, the 32 bit identifier would give a 50 % probability for collisions if the number of arrays of a certain type is 46.340. The data from all benchmarks contain at least one array type that has been seen with more than 46.340 different identities. The highest number can be found in the data from db-shootout which contains 9,608,769 identities for byte arrays.

As a consequence, we cannot rule out the possibility that more than one actual array will be assumed to be and treated to be the same array. A sign that indicates that this actually happens is that arrays' sizes sometimes seem to change. Out of all arrays in the collected data, only 0.2 % seem to change size during their lifetime, which in turn may indicate that large numbers of arrays with the same lengths are created by the benchmarks. This has impact on all analyses of access patterns since all accesses to arrays with the same id will be concatenated. If an id is first used for an array of type [T] and length 3 where each of the indices are read and written in order [r 0, w 0, r 1, w 1, r 2, w 2] and the same id is later used for another array of the type [T] and length 3 where all elements are just read in index order [r 0, r 1, r 2], the resulting access pattern will be [r 0, w 0, r 1, w 1, r 2, w 2, r 0, r 1, r 2] for these arrays assumed to be one and the same. This will prevent us from categorising these two arrays together with other arrays with the access pattern [r 0, w 0, r 1, w 1, r 2, w 2] and [r 0, r 1, r 2] respectively. This may explain the fact that some access patterns consist of combined sequences of regular traversals and what seem to be random accesses. The combined access pattern may appear to be unique although it should have been grouped together with others.

Another impact of the possibility for array identity collisions is that the number of threads accessing one array may appear higher. The number of arrays that were used by more than one thread was low, but may in fact have been even lower.

Twelve of the benchmarks were aborted once they had 10 GB of logs. This could cause imbalance in the results, but there are no patterns in the data suggesting that the aborted benchmarks will result in any particular skewing of the data. None of the metrics examined is particularly high or low for either of the groups, and high and low values are present in both groups, including the number of read-only arrays, write-only arrays, and the number of fully encoded access patterns.

**Arrays in Practice**

To examine this further, we re-ran the 12 benchmarks originally run to completion and terminated the runs after half their original run time, see Appendix D.2. Running the analysis on the new data gave similar results (within 3.3 percentage points) for shares of read-only and write-only arrays and access patterns as our original results.

## 7 Conclusions and Future Work

This paper presents a dynamic analysis of array usage and access patterns. Using our strategy for identifying sub-patterns in access patterns, we can determine that array access patterns consist of traversals to a great extent; the access pattern *repeated step increasing* together with *linear increasing* incorporates 53.8 % of all accesses. This means that the indices for more than half of the array accesses could be controlled by a for loop. Most arrays are short, most array elements are not accessed and accesses to arrays are to a great extent made using a few simple patterns.

We will continue the empirical studies of arrays by further developing the analysis method to study access patterns that are more complicated than the ones presented here. Patterns of special interest for future studiesare patterns with accesses made from two or more parallel threads. Programs of special interest to include in a future study are scientific computing programs.

We will extend the analysis to also cover the identity of objects accessing the arrays, besides the class identity. Arrays that are accessed by one single object could be embedded in that object and be omitted from tracing in garbage collection algorithms.

We will also look closer at the arrays which are accessed from more than one class to see if there is some categorisation to be made here. For example, the backing array of a hash table may be accessed by an iterator over the hash table. Another interesting category would be unpredictable access patterns to see if there is a functional relation (e.g., a hashing function) or if they are caused by truly random sources (e.g., user input or as a result of reading other input data).

Our original motivation for looking at how arrays are used is our work on *array capabilities* [1], an abstraction built on top of the array structure that offers mechanisms for splitting and merging arrays and which statically prevents data-races (e.g., due to off-by-one errors). The array capability also supports reordering of the elements stored in the underlying array to better suit the actual use e.g., by collecting all elements accessed by a certain operation in one part of the array for better cache performance. By looking at how arrays are used in real-world programs, we get further input in our continued work on array capabilities.

## A  Java Bytecode Instructions

In Java bytecode there are 20 bytecode instructions that deal with arrays that can be found in Table 6. All but 3, 4, 17 and 18 (marked with grey colour in the table) result in loading or storing values of different types in arrays.

- **Table 6** Java bytecode instructions dealing with arrays. The instructions marked by grey colour were not traced by our instrumentation.

| | | |
|---|---|---|
| 1 | aaload | Load a reference to an object from an array. |
| 2 | aastore | Store a reference to an object in an array. |
| 3 | anewarray | Create a new array of references. |
| 4 | arraylength | Getting the length of an array. |
| 5 | baload | Load a byte or boolean value from an array. |
| 6 | bastore | Store a byte or boolean value in an array. |
| 7 | caload | Load a char value from an array. |
| 8 | castore | Store a char value in an array. |
| 9 | daload | Load a double value from an array. |
| 10 | dastore | Store a double value in an array. |
| 11 | faload | Load a float value from an array. |
| 12 | fastore | Store a float value in an array. |
| 13 | iaload | Load an int value from an array. |
| 14 | iastore | Store an int value in an array. |
| 15 | laload | Load a long value from an array. |
| 16 | lastore | Store a long value in an array. |
| 17 | multinewarray | Create a new multi-dimensional array of references. |
| 18 | newarrray | Create a new array of primitive values. |
| 19 | saload | Load a short value from an array. |
| 20 | sastore | Store a short value in an array. |

## B Uninstrumented Classes

Out of the 171,481 classes used by the programs in the Renaissance benchmark suite, 168,686 (98.4%) were successfully instrumented by our tool. The identity of the uninstrumented classes and the benchmarks they belong to can be found in Table 7.

## C Arrays with Read or Write Access Only

Listing 1 shows two Java code examples that result in logging accesses to the same data in two separate arrays where one of them is write-only and the other one is read-only. Both of these examples allow the creation of arrays without explicitly specifying their storage capacity. The corresponding part of the log-file produced when the example code shown in Listing 1 was run can be found in Listing 2.

In the first example in Listing 1, on line 2 in the Java code, an IntStream is used to load the data into the program and the write operations registered in the log file (Listing 2, lines 1–4) are made for an array of int. Once the data has been read, on line 3 in the Java code, the IntStream is converted to an array (int[]).

**Arrays in Practice**

▪ **Table 7** In the middle column, the name of the jar files containing classes that were not instrumented. In the left column, the name of the benchmark they belong. In the right column the number of uninstrumented classes and the total number of classes in the jar file.

| Benchmark | jar file name | # classes (not instrumented/total) |
| --- | --- | --- |
| shared by all | commons-math3-3.6.1.jar | 3 / 1301 |
| apache-spark | hadoop-client-runtime-3.3.1.jar | 2 / 18123 |
| apache-spark | spark-catalyst_2.13-3.2.0.jar | 1 / 4025 |
| apache-spark | spark-network-common_2.13-3.2.0.jar | 1 / 1722 |
| apache-spark | spire-macros_2.13-0.17.0.jar | 50 / 50 |
| apache-spark | spire-platform_2.13-0.17.0.jar | 7 / 7 |
| apache-spark | spire-util_2.13-0.17.0.jar | 18 / 18 |
| apache-spark | spire_2.13-0.17.0.jar | 2481 / 2481 |
| database | asm-5.0.3.jar | 25 / 25 |
| database | asm-analysis-5.0.3.jar | 13 / 13 |
| database | asm-commons-5.0.3.jar | 22 / 22 |
| database | asm-tree-5.0.3.jar | 30 / 30 |
| database | asm-util-5.0.3.jar | 16 / 16 |
| neo4j | asm-9.2.jar | 30 / 30 |
| neo4j | asm-analysis-9.2.jar | 15 / 15 |
| neo4j | asm-tree-9.2.jar | 39 / 39 |
| neo4j | asm-util-9.2.jar | 25 / 25 |
| neo4j | lucene-analyzers-common-8.9.0.jar | 8 / 633 |

Listing 2 shows the log file produced when the instrumented version of the code in Listing 1 was run. Each line contains the array id, access mode, index, thread number and class id. Lines 1–4 and 10–13 show logs from write accesses made for index 0–3 in arrays of the type int and Integer respectively. The write accesses are followed by read accesses on line 5–8 and 14–17 respectively. The read accesses are made to an array with the type int in both cases and the array id is another one than the one logged with the write accesses.

▪ **Listing 1** Two examples of array creation that would result in only writes registered to one array and only reads would be registered for another array although the data is the same.

```
1  // example 1
2  IntStream stream = IntStream.of(1, 11, 111, 1111);
3  int[] a1 = stream.toArray();
4  int m = 0;
5  for (int i : a1)
6      m += i;
7
8  // example 2
9  List<Integer> lst = new ArrayList<Integer>();
10 for (Integer i : Arrays.asList(2, 22, 222, 2222))
11     lst.add(i);
12 int[] a2 = lst.stream().mapToInt(i->i).toArray();
13 int n = 0;
14 for (int i : a2)
15     n += i;
```

In the log file (Listing 2), there are no write operations registered when the data is stored anew. On lines 5–6 in the Java code (Listing 1), the elements of the array are read in a for-loop and the read operations have been registered in the log-file on lines 5–8. The identity of the array has changed between the initial write operations and the read operations and from the log file and seemingly we have two arrays where one is write-only and the other one is read-only.

■ **Listing 2** The log file produced when the instrumented version of the code in Listing 1 was run.

```
1  [I@232204a1 w 0 4 1 8 1A0C9142
2  [I@232204a1 w 1 4 1 8 1A0C9142
3  [I@232204a1 w 2 4 1 8 1A0C9142
4  [I@232204a1 w 3 4 1 8 1A0C9142
5  [I@5cad8086 r 0 4 1 13 1A0C9142
6  [I@5cad8086 r 1 4 1 13 1A0C9142
7  [I@5cad8086 r 2 4 1 13 1A0C9142
8  [I@5cad8086 r 3 4 1 13 1A0C9142
9
10 [Ljava.lang.Integer;@6e0be858 w 0 4 1 18 1A0C9142
11 [Ljava.lang.Integer;@6e0be858 w 1 4 1 18 1A0C9142
12 [Ljava.lang.Integer;@6e0be858 w 2 4 1 18 1A0C9142
13 [Ljava.lang.Integer;@6e0be858 w 3 4 1 18 1A0C9142
14 [I@60e53b93 r 0 4 1 24 1A0C9142
15 [I@60e53b93 r 1 4 1 24 1A0C9142
16 [I@60e53b93 r 2 4 1 24 1A0C9142
17 [I@60e53b93 r 3 4 1 24 1A0C9142
```

A similar situation occurs in the second example where the data is originally stored in an ArrayList<Integer> (on lines 9–11 in Listing 1). The ArrayList is then, on line 12 in the Java code, transformed and stored as an array, and the elements of the array are read on lines 14–15. The creation of the ArrayList has been registered as write operations in the log file (Listing 2, lines 10–13). These write operations are made for an array of Integer. Once again the log file contains no trace of the write operations made when the data is stored again, but the read operations have been logged (Listing 2, lines 14–17). Just as in the first example the identity of the array is another when the read operations are registered and in this case the type is also different, that is int[] instead of Integer[].

## D  Supplementary Experiments

To examine the validity of our results, we conducted two supplementary experiments. The first one to examine the possible inconsistency caused by the inclusion of Scala standard library classes while excluding Java standard library classes and the second one to examine the effect of aborting 13 out of the 25 benchmarks after collecting at least 10 GB of data.

### D.1  Scala Standard Library

To examine the possible inconsistency caused by the inclusion of Scala standard library classes while excluding Java standard library classes, we separated all arrays that were used from classes within the Scala standard library and re-run our analyses on these arrays. This analysis collected 34,060,319 arrays, that is 36 % of the 93,647,508

**Arrays in Practice**

arrays in the original analysis. Most of the benchmarks that produced the highest numbers of arrays are part of Apache Spark, and implemented in Scala.

The Scala only access pattern analysis did produce similar results as did the analysis of the entire data.

Access pattern identification from article:

|         | Completely encoded | Partially | Completely unencoded |
|---------|--------------------|-----------|----------------------|
| Round 1 | 69.8%              | 19.3%     | 10.9%                |
| Round 2 | 81.4%              | 11.4%     | 7.2%                 |

Access pattern identification from Scala's standard library only:

|         | Completely encoded | Partially | Completely unencoded |
|---------|--------------------|-----------|----------------------|
| Round 1 | 71.2%              | 17.0%     | 11.8%                |
| Round 2 | 80.3%              | 9.0%      | 6.5%                 |

For the read-only and write-only arrays, the Scala only analysis produced a higher average for read-only arrays and a lower average for the write-only arrays than the original analysis. Three out of four benchmarks that had over 90% read-only arrays in the original analysis were Scala benchmarks, while all three benchmarks that had 100% write-only arrays were Java benchmarks.

Read-only and write-only from article:

|           | r-o   | w-o    |
|-----------|-------|--------|
| Avg share | 13.8% | 41.2%  |
| Max share | 76.1% | 100.0% |
| Min share | 0.0%  | 0.9%   |

Read-only and write-only from Scala's standard library only:

|           | r-o   | w-o   |
|-----------|-------|-------|
| Avg share | 26.5% | 17.6% |
| Max share | 55.6% | 60.5% |
| Min share | 0.0%  | 0.0%  |

For the three benchmarks Scala benchmarks that had the highest share of read-only arrays in the original analysis; page-rank (97.8%), als (93.0%) and naive-bayes (91.5%), the share of read-only arrays was much lower when only the Scala standard library part of them was analysed; page-rank (6,4%), als (29,1%) and naive-bayes (23,55%). This indicates that the high number of read-only arrays is not specifically a characteristic of the Scala standard library.

### D.2 Aborted Benchmarks

Since 13 out of the 25 benchmarks were aborted after collecting at least 10 GB of data, there may be a risk of skewing of the data.

There are no patterns in the data from the original analysis that suggests that aborted benchmarks will result in any particular skewing of the data. None of the metrics we have examined is particularly high or low for either of the groups, but high and low values are present in both groups. Some examples are the presence of arrays that have only been read from, arrays that have only been written to, and the number of arrays that could be matched completely with access patterns. There are high and low values for all of these both in the group of benchmarks that were aborted by us and in the group that were allowed to run to completion.

To examine this further, we run the 12 benchmarks that originally were run to completion and aborted the runs after half the time it took for them to complete. Running the analysis on the new data gave similar results as the results presented in the article.

Read-only and write-only from article:

|           | r-o    | w-o     |
|-----------|--------|---------|
| Avg share | 13.8 % | 41.2 %  |
| Max share | 76.1 % | 100.0 % |
| Min share | 0.0 %  | 0.9 %   |

Read-only and write-only from experiment:

|           | r-o    | w-o     |
|-----------|--------|---------|
| Avg share | 13.5 % | 40.1 %  |
| Max share | 69.5 % | 100.0 % |
| Min share | 0.0 %  | 0.0 %   |

Access pattern identification from article:

|         | Completely encoded | Partially | Completely unencoded |
|---------|--------------------|-----------|----------------------|
| Round 1 | 69.8 %             | 19.3 %    | 10.9 %               |
| Round 2 | 81.4 %             | 11.4 %    | 7.2 %                |

Access pattern identification from experiment:

|         | Completely encoded | Partially | Completely unencoded |
|---------|--------------------|-----------|----------------------|
| Round 1 | 71.9 %             | 22.6 %    | 11.3 %               |
| Round 2 | 79.0 %             | 12.7 %    | 8.3 %                |

The average share of read-only and write-only in the supplementary experiment analysis were within 1.1 % of the shares in the original analysis.

The average share of identified access patterns in the supplementary experiment analysis were all within 3.3 % of the shares in the original analysis.

## E The Renaissance Benchmark Suite

Table 8 shows an overview of the benchmarks included in the Renaissance suite. In the rightmost column a J indicates that the program is a Java program while an S indicates that the program is a Scala program.

**Arrays in Practice**

▪ **Table 8** The benchmarks included in the Renaissance benchmark suite are organised in groups. Information from https://renaissance.dev/docs (Last accessed 2024-02-21)

| group | benchmark | description | lang |
|---|---|---|---|
| apache-spark | als | runs ALS algorithm from spark.ml | S |
| | chi-square | runs chi-square test from spark.ml | S |
| | dec-tree | runs Random Forest algorithm from spark.ml | S |
| | gauss-mix | computes Gaussian mixture model using expectation-maximization. | S |
| | log-regression | runs logistic regression algorithm from spark.ml | S |
| | movie-lens | recommends movies using the ALS algorithm | S |
| | naive-bayes | runs multinomial naive Bayes algorithm from spark.ml | S |
| | page-rank | runs a number of PageRank iterations, using RDDs | S |
| functional | future-genetic | runs a genetic algorithm using the Jenetics library and futures | J |
| | mnemonics | solves the phone mnemonics problem using JDK streams | J |
| | par-mnemonics | solves the phone mnemonics problem using parallel JDK streams | J |
| | rx-scrabble | solves the Scrabble puzzle using the Rx streams | J |
| | scrabble | solves the Scrabble puzzle using JDK Streams | J |
| concurrency | akka-uct | runs unbalanced cobwebbed tree actor workload in Akka | S |
| | fj-kmeans | runs k-means algorithm using the fork/join framework | J |
| | reactors | runs benchmarks inspired by the Savina micro benchmark workloads in a sequence on Reactors.IO | S |
| database | db-shootout | runs a shootout test using several in-memory databases | J |
| | neo4j-analytics | runs Neo4J graph queries against a movie database | S |
| scala | dotty | runs the Dotty compiler on a set of source code files | S |
| | philosophers | solves a variant of the dining philosophers problem using ScalaSTM | S |
| | scala-doku | solves Sudoku Puzzles using Scala collections | S |
| | scala-kmeans | runs the K-Means algorithm using Scala collections | S |
| | scala-stm-bench7 | runs the stmbench7 benchmark using ScalaSTM | S |
| web | finagle-chirper | simulates a microblogging service using Twitter Finagle | S |
| | finagle-http | sends many small Finagle HTTP requests to a Finagle HTTP server and awaits response | S |
| dummy | | no benchmarks from this group were included in the study | |

## About the authors

**Beatrice Åkerblom** is a PhD student at Stockholm University. Her research interests are programming language design and typability. Contact her at beatrice@dsv.su.se.
https://orcid.org/0009-0005-2855-136X

**Elias Castegren** is an assistant professor at Uppsala University. His research interests are programming language semantics, type systems and concurrency. Contact him at elias.castegren@it.uu.se.
https://orcid.org/0000-0003-4918-6582